# Scaling mean annual peak flow scaling with upstream basin area

Behzad Ghanbarian

Department of Earth and Environmental Sciences, University of Texas at Arlington, Arlington TX 76019, United States

Department of Civil Engineering, University of Texas at Arlington, Arlington TX 76019, United States

Division of Data Science, College of Science, University of Texas at Arlington, Arlington TX 76019, United States

\* Corresponding author's email address: ghanbarianb@uta.edu

**Abstract**

Understanding how annual peak flow, $Q_p$, relates to upstream basin area, $A$, and their scaling have been one of the challenges in surface hydrology. Although a power-law scaling relationship (i.e., $Q_p \propto A^\alpha$) has been widely applied in the literature, it is purely empirical, and due to its empiricism the interpretation of its exponent, $\alpha$, and its variations from one basin to another is not clear. In the literature, different values of $\alpha$ have been reported for various datasets and drainage basins of different areas. Invoking concepts of percolation theory as well as self-affinity, we derived universal and non-universal scaling laws to theoretically link $Q_p$ to $A$. In the universal scaling, we related the exponent $\alpha$ to the fractal dimensionality of percolation, $D_x$ (i.e., $\alpha = 1 + 0.5(1 - D_x)$). In the non-universal scaling, in addition to $D_x$, the exponent $\alpha$ was related to the Hurst exponent, $H$, characterizing the boundaries of the drainage basin (i.e., $\alpha = 1 + (1 - D_x)/(1 + H)$). The $D_x$ depends on the dimensionality of the drainage system (e.g., two or three dimensions) and percolation class (e.g., random or invasion percolation). We demonstrated that the theoretical universal and non-universal bounds were in well agreement with experimental ranges of $\alpha$ reported in the literature. More importantly, our theoretical framework revealed that greater $\alpha$ values are theoretically expected when basins are more quasi two-dimensional, while smaller values when basins are mainly quasi three-dimensional. This is well consistent with the experimental data. We attributed it to the fact that small basins most probably display quasi-two-dimensional topography, while large basins quasi-three-dimensional one.

**Keywords:** Drainage basin, Mean annual peak flow, Fractal dimension, Optimal path, Shortest path

## 1. Introduction

Despite recent progress, one of the ongoing challenges in hydrologic sciences has been scaling. In the National Research Council report "Challenges and Opportunities in the Hydrologic Sciences" (National Research Council, 2012), it is stated that challenges about how hydrologic processes at smaller scales interact with those at larger scales still exist. Experimental data cannot be measured at all ranges of temporal and spatial scales (e.g., within a watershed). In addition to that, physically-based hydrologic models need extensive data to be validated and/or calibrated. Thus, predictive scaling analyses are not only theoretically interesting but also practically important (National Research Council, 2012).

It is now more than thirty years since James A. Smith in 1992 published his article entitled "Representation of basin scale in flood peak distributions". Smith (1992) analyzed annual flood peaks from 104 stations in the central Appalachian region of Maryland and Virginia. In his study, the basin drainage area spanned nearly five orders of magnitude from nearly 0.78 to around 26,100 km². Based on collected experimental data, Smith (1992) showed that the volumetric mean annual peak flow, $Q_p$ $[L^3 T^{-1}]$, scaled with the upstream basin area, $A$ $[L^2]$, through the following power law:

$$Q_p \propto A^\alpha \qquad (1)$$

where $\alpha$ is an empirical exponent whose value depends on the topologic structure of river networks (Gupta and Waymire, 1998). Different values for the exponent $\alpha$ reported in the literature are summarized in Table 1. For instance, Smith (1992) found $\alpha$ in the range of 0.65 and 0.69 with an average value of 0.67.

Although widely applied, the physical origin of Eq. (1) has remained unsolved, and theoretical insights into the exponent $\alpha$ unavailable. In the literature, attempts were made to explain the basin area dependence of $Q_p$. For example, Sivapalan et al. (2002) proposed a storage model consisting of two component processes i.e., runoff generation on hillslopes and flow routing in a channel network. For the sake of simplicity, Sivapalan et al. (2002) assumed that those stores were linear and arranged in series (see their Fig. 1) and proposed the following relationship between $Q_p$ and $A$

$$Q_p \propto A \left[ 1 - exp\left(-\frac{t_r}{t_c}\right) \right] \qquad (2)$$

where $t_r$ is the residence time, and $t_c$ is the characteristic time. The value of $t_c$ in a basin can be estimated from $d_t/u$ where $d_t$ is the traveling distance and $u$ is the average traveling velocity. Eq. (2) results in a linear trend between $Q_p$ and $A$ when $t_c < t_r$, while a nonlinear and power law-like trend, similar to Eq. (1), when $t_c > t_r$ (see Fig. 2c in Sivapalan et al. (2002)).

Eq. (1) is purely empirical, and due to its empiricism the interpretation of its exponent, $\alpha$, and its variations from one basin to another is not clear. In addition, it does not provide deep insight about the relationship between the mean annual peak flow and upstream basin area. More specifically, it is not clear how the empirical exponent $\alpha$ in Eq. (1) can be determined prior to collecting experimental data. Therefore, our objectives are to: (1) establish a theoretical relationship between the mean annual peak flow, $Q_p$, and the upstream basin area, $A$, and (2) provide a physical interpretation for the exponent in the empirical power-law equation widely used by various researchers in the literature.

In what follows, we first briefly introduce percolation theory and its concepts. We then invoke concepts of percolation theory and self-affinity to theoretically link the mean annual peak flow, $Q_p$, to the upstream basin area, $A$. More specifically, we derive universal and non-universal scaling laws and demonstrate that the exponent $\alpha$ in Eq. (1) depends on percolation fractal dimension and/or self-affine properties of basin boundaries.

## 2. Percolation theory

Percolation theory, introduced in its present form by Broadbent and Hammersley (1957), is a theoretical approach from statistical physics for quantifying transport properties in complex systems. It provides a promising framework to study interconnectivity and its effects on transport properties in heterogeneous systems and complex networks.

Percolation theory exists in three main forms: bond, site, and continuum percolation. For the sake of simplicity, we briefly introduce bond percolation here. The interested reader is referred to Stauffer and Aharony (1994) for other forms. In bond percolation, bonds, randomly distributed segments open to transport, act as conducting elements. Fig. 1 schematically shows an example of bond percolation using a finite square network. In Fig. 1a, the occupation probability $p$ is equal to 0.33, which is below the threshold of an infinite square network whose $p_c$ is 0.5. In Fig. 1a, although there exist several finite clusters with different sizes, a sample-spanning cluster, connecting two opposing faces of the network, has not formed yet. As the occupation probability $p$ increases toward 0.5 (the critical occupation probability), many individual clusters become connected to each other, making a large interconnected cluster. Figure 1b presents the square network with $p = 0.55$ ($p > p_c = 0.5$), where a large cluster connects the top side of the square network to its bottom side, and the left to the right, which means that percolation across the network has occurred. Although we introduced random bond percolation here, there exist others types, such as invasion, directed and gradient, used to model different phenomena (Hunt et al., 2014; Sahimi, 2023).

Within the percolation theory framework, there exist power-law scaling laws to explain transport phenomena in complex networks. For instance, Lee et al. (1999) studied traveling times for tracer particles and modeling flow driven by a pressure difference between two points separated by Euclidean distance $x$ (Fig. 2). They demonstrated that the most probable traveling time, $t$, of particles on a percolation cluster is proportional to $x^{D_x}$. Therefore,

$$x \propto t^{\frac{1}{D_x}} \tag{3}$$

where $D_x$ is the relevant fractal dimension and the subscript $x$ stands for either $b$, denoting the backbone cluster spanning the network, $opt$, denoting the optimal path, or $min$, denoting the shortest path. We present different values of $D_x$ for various percolation classes (e.g., for random and invasion percolation) and system dimensions in Table 2.

Eq. (3) and its subsequent derivations have been successfully evaluated in various areas, such as describing the weathering rate of porous media (Hunt, 2015), soil depth and soil production (Hunt and Ghanbarian, 2016), surface flow (Ghanbarian et al., 2018) and geochemical reaction rates (Hunt et al., 2015). Eq. (3) is also similar to the Boltzmann scaling (i.e., $x \propto \sqrt{t}$) in which the exponent linking $x$ to $t$ is equal to 0.5, corresponding to $D_x = 2$ in Eq. (3), regardless of the system dimensionality.

In Fig. 3, we schematically show a basin with the drainage area $A$ and mainstream length $L$. In this figure, as shown, $L_s$ represents the straight-line length, the Euclidean distance connecting the two ends of the longest stream in the basin. By taking the first derivative of Eq. (3), one finds the following relationship to explain the average velocity, $v$, as a function of the traveling time, $t$, and/or straight-line length, $L_s$,

$$v = \frac{dL_s}{dt} \propto t^{\frac{1-D_x}{D_x}} \propto L_s^{1-D_x} \tag{4}$$

Volumetric mean annual peak flow, $Q_p$, is the average of peak flows within a basin over a year. Peak flows in a basin correspond to either intermediate to heavy rainfalls occurring all over the drainage area or heavy precipitations over a small area within the basin. Here, we assume that the former is more typical than the latter. Accordingly, multiplying both sides of Eq. (4) by the upstream basin area, $A$, gives

$$Q_p \propto vA \propto At^{\frac{1-D_x}{D_x}} \propto AL_s^{1-D_x} \tag{5}$$

In the next sections, by invoking concepts from percolation theory we derive scaling laws based on Eq. (5) to shed light on the empirical exponent $\alpha$ in Eq. (1). We demonstrate that $\alpha$ is related to fractal properties that characterize transport processes within a basin and its self-affine characteristics.

## 3. Universal scaling law

Montgomery and Dietrich (1992) showed that the straight-line distance, $L_s$, depends on the area $A$ over seven orders of magnitude of length scale as follows

$$L_s \propto A^{0.5} \tag{6}$$

Replacing $L_s$ in Eq. (5) with $A^{0.5}$ from Eq. (6) yields

$$Q_p \propto A^{1+0.5(1-D_x)} \tag{7}$$

Comparing Eq. (7) with Eq. (1) clearly shows that the empirical exponent $\alpha$ can be theoretically interpreted within the percolation theory framework and, thus, has some physical meaning, as we discuss in the following.

Different values of $D_x$ for various percolation classes (e.g., for random and invasion percolation) and system dimensions are summarized in Table 2. As can be seen, in two dimensions $1.13 \leq D_x \leq 1.643$ and, correspondingly, $0.68 \leq \alpha = 1 + 0.5(1 - D_x) \leq 0.94$. Interestingly, the upper bound 0.94 is consistent with the linear trend produced by the Sivapalan et al. (2002) model, Eq. (2), when $t_c < t_r$. In three dimensions, we have $1.376 \leq D_x \leq 1.87$ corresponding to $0.57 \leq \alpha = 1 + 0.5(1 - D_x) \leq$ 0.81. These wide ranges are well consistent with those experimentally reported in the literature and summarized in Table 1.

It can be deduced from Table 1 that greater $\alpha$ values belong to smaller upstream basin areas. Given that the exponent $1 + 0.5(1 - D_x)$ returns greater $\alpha$ values in two dimensions (Table 2), it may be inferred that river flow in smaller basins could be quasi two-dimensional. The smaller the basin, the lower the probability that basin elevation and its topography greatly vary within the basin. This means that flow most probably would be mainly restricted to a planar surface in two dimensions.

For the Boltzmann scaling corresponding to $D_x = 2$ in Eq. (3), the exponent $1 + 0.5(1 - D_x)$ would be equal to 0.5 (both in two and three dimensions), close to $\alpha = 0.52$ reported by Di Lazzaro and Volpi (2011) for the Tiber River basin, central Italy (see Table 1). It seems that the Boltzmann scaling provides the lower bound of the exponent $\alpha$. However, it does return the same exponent for two and three dimensions, in contrast to percolation theory (Table 2).

## 4. Non-universal scaling law

For basins with self-affine boundaries, Tricot (1995) proposed that

$$L_s \propto A^{1/(1+H)} \tag{8}$$

where $H$ is the Hurst exponent that characterizes the self-affinity of the basin boundaries ($0 \leq H \leq 1$). $H < 1$ indicates that the basin boundaries are self-affine, while $H = 1$ corresponds to self-similarity (Rodriguez-Iturbe and Rinaldo, 1997).

Combining Eq. (8) with Eq. (5) results in

$$Q_p \propto A^{1+\frac{1-D_x}{1+H}} \tag{9}$$

which is similar in form to Eq. (7). However, $H$ in Eq. (9) is non-universal meaning that its value may vary from one basin to another. Comparing Eq. (9) with Eq. (1) demonstrates that the empirical exponent $\alpha$ can be related to the physically meaningful percolation fractal dimension, $D_x$, and the Hurst exponent characterizing the self-affinity of basin boundaries.

If one assumes that the traveling time follows the shortest path concepts from percolation theory, then the fractal dimension $D_x$ in Eq. (8) should change to $D_{min}$, the shortest path (or minimum) fractal dimension, whose value is 1.13 or 1.374 in two or three dimensions, respectively (Porto et al., 1997). Fig. 4 shows the exponent $\alpha = 1 + (1 - D_x)/(1 + H)$ versus the Hurst exponent $H$ ranging from 0 to 1. In two dimensions, we found $0.87 \leq \alpha = 1 + (1 - D_x)/(1 + H) \leq 0.94$ (blue dashed line) for $0 \leq H \leq 1$. In three dimensions, results, presented in Fig. 4, showed that as the $H$ increases from 0 to 1, the value of $\alpha$ increases from 0.63 to 0.82 ($0.63 \leq 1 + (1 - D_x)/(1 + H) \leq 0.82$) (red dashed line). This range is consistent with the experimental one reported in Table 1.

The most interesting path across a basin may not be the shortest (Hunt, 2016). The optimal path is the most energetically favorable and least-cost path whose fractal dimension, $D_{opt}$, is 1.21 in two and 1.43 in three dimensions (Table 2). Results presented in Fig. 4 indicate that as the $H$ increases from 0 to 1, the exponent $\alpha$ increases from 0.79 to 0.9 in two (blue solid line) and from 0.57 to 0.79 in three (red solid line) dimensions, in accord with the experimental range presented in Table 1.

For random percolation, we found lower $\alpha$ values compared to the shortest and optimal paths (Fig. 4). In random percolation, $D_x = D_b$ and equal to 1.64 in two and 1.87 in three dimensions (Table 2). We found $0.36 \leq \alpha = 1 + (1 - D_x)/(1 + H) \leq 0.68$ in two dimensions and $0.13 \leq \alpha = 1 + (1 - D_x)/(1 + H) \leq 0.57$ in three dimensions.

## 5. Discussion

In our study, we developed universal and non-universal scaling laws to theoretically explain the power-law relationship experimentally observed between the mean annual peak flow and the upstream basin area. In fact, our non-universal scaling laws, Eqs. (8) and (9), reduce to our universal derivations, Eqs. (6) and (7), when one sets $H = 1$. Ijjasz-Vasquez et al. (1994) analyzed properties of boundaries of eight drainage basins, determined the Hurst exponent and found $0.74 \leq H \leq 0.79$ with an average value of 0.755. This average value in combination with $D_{opt} = 1.43$ (three dimensions) results in $\alpha = 1 + (1 - D_x)/(1 + H) = 0.75$ in Eq. (8), which is not greatly different from $\alpha = 0.70$ reported by Knisel (1980) or 0.67 observed by Smith (1992) (Table 1). In another study, Rigon et al. (1996) analyzed thirteen basins with area between 200 and 2000 km$^2$ and determined $0.75 \leq H \leq 1.01$ (see their Table 2). The average Hurst exponent equal to 0.93 in Rigon et al. (1996) corresponds to $\alpha = 1 + (1 - D_x)/(1 + H) = 0.78$ assuming that $D_x = D_{opt} = 1.43$. This $\alpha$ value is close to 0.83 reported by Ogden and Dawdy (2003) for basins with areas ranged from 0.172 to 21 km$^2$. Let us assume that basins analyzed by Ogden and Dawdy (2003) were quasi two-dimensional. Under such a condition, one may set $D_{opt} = 1.32$, the average of 1.21 (2D optimal path fractal dimension) and 1.43 (3D optimal path fractal dimension). Then, using $H = 0.93$, one has $\alpha = 0.83$, observed experimentally by Ogden and Dawdy (2003).

One should note that theoretically the Hurst exponent should not exceed 1, although $H > 1$ has been reported in the literature. For instance, Nikora (1994) analyzed twenty five rives from around the world and found $H > 1$ for five of them. He attributed that to errors in the calculation of $H$ as well as possible anomalous geological-geomorphological conditions.

For a fractal path the mainstream length, $L$, would depend on the straight-line length, $L_s$, as follows (Mandelbrot, 1982; Wheatcraft and Tyler, 1988)

$$L \propto L_s^{D_r} \tag{10}$$

in which $D_r$ is the fractal dimensional characterizing mainstream meandering properties. Hunt (2016) argued that mapping a river network onto a planar surface makes the topology of stream connections two-dimensional and, thus, 2D fractal dimensions would be more relevant in the river network analysis. Based on such an argument, he proposed that $D_r$ should range between 1.13 ($D_{min}$ in two dimensions) to 1.21 ($D_{opt}$ in two dimensions), with an average value of 1.17. Then, by invoking the Montgomery and Dietrich (1992) relationship, $L_s \propto A^{0.5}$, Hunt (2016) found a theoretical power law between $L$ and $A$ with an exponent 0.585 (i.e., $L \propto A^{0.585}$), close to 0.6 in Hack's law (Hack, 1957). Interesting, by analyzing thirty four basins from West Virginia Rodriguez-Iturbe and Rinaldo (1997) experimentally found $D_r$ in Eq. (10) to be 1.11 (see their Fig. 2.57), close to $D_{min} = 1.13$, the shortest path fractal dimension in two dimensions. In another study, Rigon et al. (1996) found $D_r = 1.15$, near the average 1.17 proposed by Hunt (2016). Nikora et al. (1996), however, found a slightly smaller value i.e., $D_r = 1.02$ for the Hutt River basin in New Zealand.

## 6. Conclusions

In this study, we presented theoretical insights from percolation theory for the exponent $\alpha$ in the empirical scaling law, Eq. (1), linking the volumetric mean annual peak flow, $Q_p$, to the upstream basin area, $A$. We proposed universal and non-universal scaling laws (i.e., Eqs. (7) and (9)) and demonstrated that the exponent $\alpha$ depends on the percolation fractal dimension and/or self-affine properties of the basin boundaries. By universal scaling law, we mean the exponent $\alpha$ only depends on the universal percolation exponents and dimensionality of the system. In the non-universal scaling law, in addition to universal percolation exponents, the $\alpha$ is a function of the Hurst exponent, ranging between 0 and 1, whose value might vary from one basin to another. For the universal scaling law, we $0.68 \leq \alpha \leq 0.94$ in two dimensions and $0.57 \leq \alpha \leq 0.81$ in three dimensions. These wide ranges are well consistent with those experimentally reported in the literature. For the non-universal scaling law, in two dimensions we found that as the Hurst exponent $H$ increased from 0 to 1 the exponent $\alpha$ increased from 0.87 to 0.94 for the shortest path and from 0.79 to 0.90 for the optimal path. In three dimensions, however, $\alpha$ increased from 0.63 to 0.82 for the shortest path and from 0.57 to 0.79 for the optimal path. These wide ranges are well consistent with those experimentally reported in the literature. Insights from percolation theory revealed that one should expect the exponent $\alpha$ to be greater in two-dimensional basins, while smaller in three-dimensional ones.

**Acknowledgement**

The author is grateful to Murugesu Sivapalan, University of Illinois at Urbana-Champaign, for fruitful discussion on the Sivapalan et al. model, and to the University of Texas at Arlington for financial supports through faculty startup fund. This modest contribution is dedicated to the memory of the late Professors Ignacio Rodriguez-Iturbe, Princeton University, and Vijay K. Gupta, University of Colorado at Boulder, for their seminal contributions to surface hydrology and river networks.

Table 1. The value of the exponent $\alpha$ reported in the literature. Some data were adapted from Liu et al. (2017).

| Reference | Region | $\alpha$ | Upstream basin area (km$^2$) |
|---|---|---|---|
| (Goodrich et al., 1997) | Walnut Gulch, AZ, USA | 0.85 | 0.00183-1 |
| (Ogden & Dawdy, 2003) | Goodwin Creek basin, MS, USA | 0.83 | 0.172-21 |
| (Knisel, 1980) | USA | 0.70 | 0.71-62 |
| (Smith, 1992) | Central Appalachian | 0.67 | 0.78-26,100 |
| (Fu et al., 2008) | Chabagou, Zizhou, Shaanxi, China | 0.59 | 0.21-96 |
| (Benson, 1964) | Texas, New Mexico, etc. USA | 0.59 | 2.5-91,000 |
| (Goodrich et al., 1997) | Walnut Gulch, AZ, USA | 0.55 | 1-149 |
| (Gupta et al., 2010) | Iowa river basin, IA, USA | 0.54 | 6.6-32,374 |
| (Di Lazzaro & Volpi, 2011) | Tiber River, Central Italy | 0.52 | 218-4116 |
| (Smith, 1992) | Central Appalachian | 0.67 | 0.78-26,100 |

Table 2. Fractal dimensionality (Sheppard et al., 1999) and the corresponding universal $\alpha$ values for various percolation classes in two and three dimensions. RP represents the random percolation and TIP the trapping invasion percolation.

| Dimension | Percolation class | $D_x$ | $\alpha = 1 + 0.5(1 - D_x)$ |
|---|---|---|---|
| | **Backbone** | | |
| 2 | RP | 1.643 | 0.68 |
| 2 | Site TIP | 1.217 | 0.89 |
| 2 | Bond TIP | 1.217 | 0.89 |
| | **Optimal path** | | |
| 2 | Optimal path | 1.210 | 0.90 |
| | **Shortest path** | | |
| 2 | Shortest path | 1.130 | 0.94 |
| | **Backbone** | | |
| 3 | RP | 1.870 | 0.57 |
| 3 | Site TIP | 1.861 | 0.57 |
| 3 | Bond TIP | 1.458 | 0.77 |
| | **Optimal path** | | |
| 3 | Optimal path | 1.420 | 0.79 |
| | **Shortest path** | | |
| 3 | Shortest path | 1.376 | 0.81 |

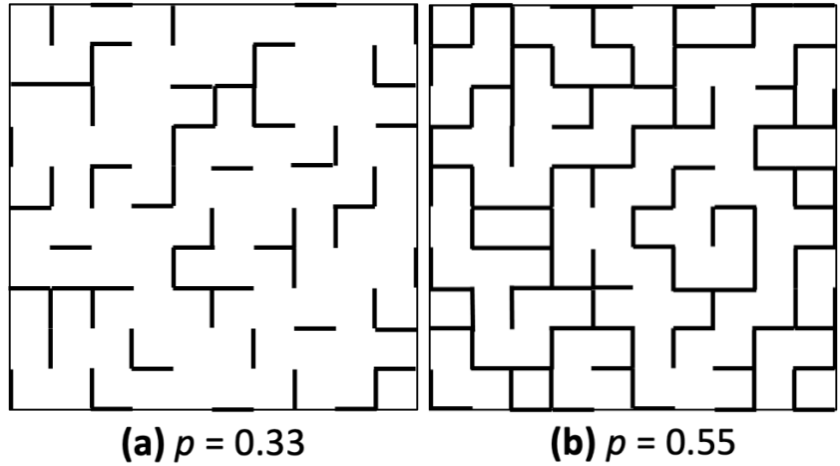

Figure 1. Scheme of bond percolation on a square lattice for two different occupation probabilities $p$: (a) $p = 1/3$ ($p < p_c$), and (b) $p = 0.55$ ($p > p_c$). $p_c = 0.5$ for bond percolation on an infinitely-large square lattice.

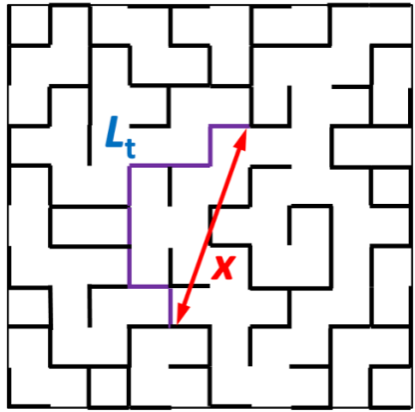

Figure 2. One possible traveling path from one point on a schematic square network to another, indicated by $L_t$ in purple. The corresponding Eulidean distance between these two points is $x$ shown in red. Rsults of Lee et al. (1999) showed that the Euclidean distance is related to the most probable traveling time via Eq. (2).

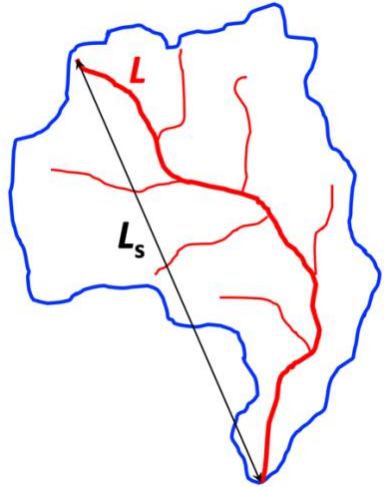

Figure 3. A schematic basin with drainage area, $A$, and mainstream length, $L$, shown by a thick red line. $L_s$ represents the straight-line length, the Euclidean distance connecting the two ends of the longest stream in the basin.

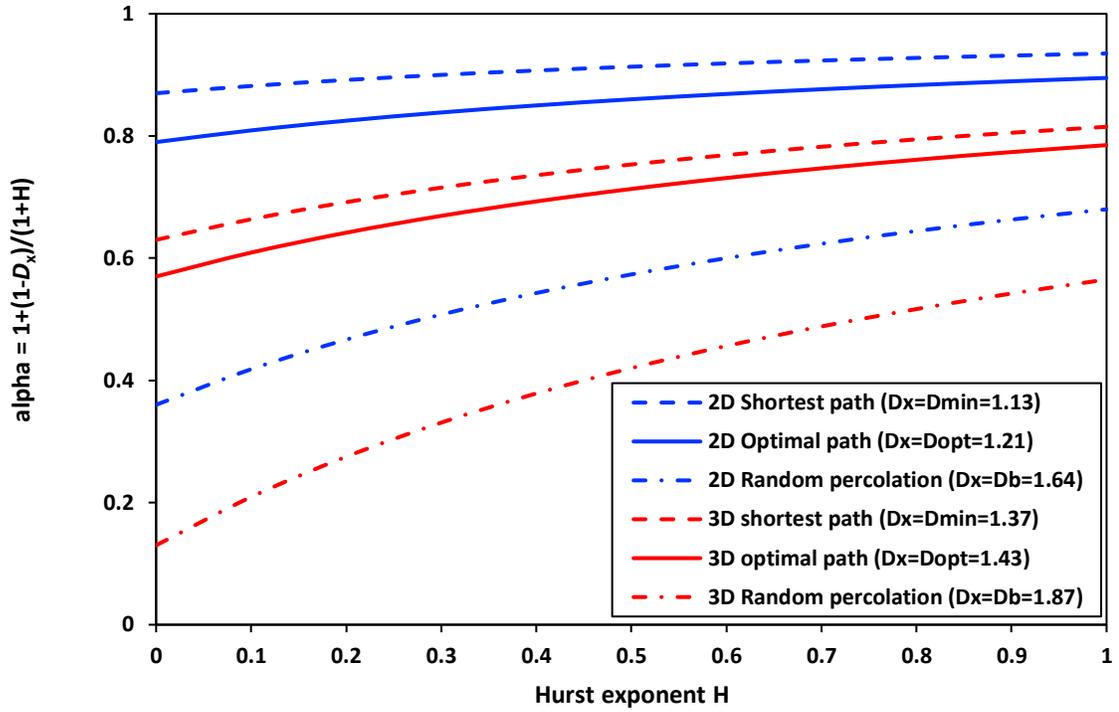

Figure 4. Non-universal exponent $1 + (1 - D_x)/(1 + H)$ versus Hurst exponent ($0 \leq H \leq 1$) for shortest path ($D_x = D_{min}$; $D_{min}$ is the shortest path fractal dimension), optimal path ($D_x = D_{opt}$; $D_{opt}$ is the optimal path fractal dimension) and random percolation ($D_x = D_b$; $D_b$ is the backbone fractal dimension). $D_{min} = 1.13$, $D_{opt} = 1.21$ and $D_b = 1.64$ in two dimensions, and $D_{min} = 1.37$, $D_{opt} = 1.43$ and $D_b = 1.87$ in three dimensions (Table 2).